\documentstyle[11pt,newpasp,twoside,epsf]{article}

\begin{document}

\title{Monte Carlo Simulations of Dense Galactic Nuclei.}

\author{Marc Freitag}
\affil{Observatoire de Gen{\`e}ve, CH-1290~Sauverny, Switzerland}
\author{Willy Benz}
\affil{Physikalisches Institut, Universit{\"a}t Bern, Sidlerstrasse 5,
  CH-3012~Bern, Switzerland}


\section{Introduction}
The presence of massive black holes (BHs) in the center of many
galaxies appears as an inescapable conclusion of recent
high-resolution observations. This fact revives much interest in the
study of the joint evolution of the central BH and its surrounding
stellar cluster. In particular, in systems with high stellar
densities, bright accretion flares are bound to occur when stars are
destroyed by the BH's tidal forces or through collisions with other
stars.

In the past few years, we have written a new code to simulate the
evolution of the central star cluster (consisting of $10^6$--$10^9$
stars) over $10^9$--$10^{10}$\,years. As disruptive events are more
likely to occur in dense nuclei, we focused on such systems and
devised a code that includes the most relevant physical processes,
i.e., 2--body relaxation, tidal disruptions, stellar collisions,
\ldots(stellar evolution to be added in the next development stage).
In our Monte Carlo (MC) scheme, based on the work by H{\'e}non (1973),
the cluster's self-gravity as well as the BH's growth are naturally
coped for and no restriction applies to the stellar mass spectrum or the
velocity distribution. The principal limitations of the method stem
from its most powerful simplifying assumptions, namely that the stellar
system is relaxed and that it obeys strict spherical symmetry. A
complete description of this code is to be found in Freitag \& Benz
(2000a). It has recently been complemented with a module that uses the
results of a large set of SPH (Smoothed Particle Hydrodynamics)
simulations (Freitag \& Benz 2000b) to implement stellar collisions
with unpreceded realism.
 
\section{Cluster Simulations with Stellar Collisions}
Collisions have long been envisioned to play a key dynamical role in
the evolution of dense galactic nuclei. Indeed, the early simulation
works that included these events, showed that they can not only feed
the central black hole with important amount of stellar
gas but that they would also imprint the structure of the stellar
cluster.  Most noticeably, in these computations, the disruption of
stars in the central regions prevent the formation of a steep
$R^{-\alpha}$ density cusp with $\alpha\simeq 1.75$ but yielded a mild
$\alpha\simeq 0.5$.  Unfortunately, the way collisions were included
in these previous papers cannot be claimed to be realistic. First, very
simplistic recipes were used to determine the outcome of collisions;
mainly the assumption of complete disruption or some semi-analytical
treatment similar to the one invented by Spitzer \& Saslaw (1966).
Furthermore, numerical schemes that perform a direct integration of
the Fokker-Planck equation impose a fixed (relatively coarse)
discretization of the mass spectrum. Consequently, even when partial
disruptions or mergers are accounted for, the resulting stars are
distributed over the existing mass classes in a quite unphysical way.

In the MC code, each particle represents a set of stars sharing the
same physical properties.  This particle-based approach, very similar
to the $N$-body philosophy, allows arbitrary stellar masses and
orbital properties. Thus, any prescription can be used to set the
outcome of stellar collisions.  To take the best advantage of this
feature, we have computed a huge number ($\simeq 14000$) of SPH
simulations of collisions between MS stars. The outcome of any given
collision is determined by interpolation into this database (Freitag
\& Benz 2000b).

\begin{figure}[t!]
  \plotone{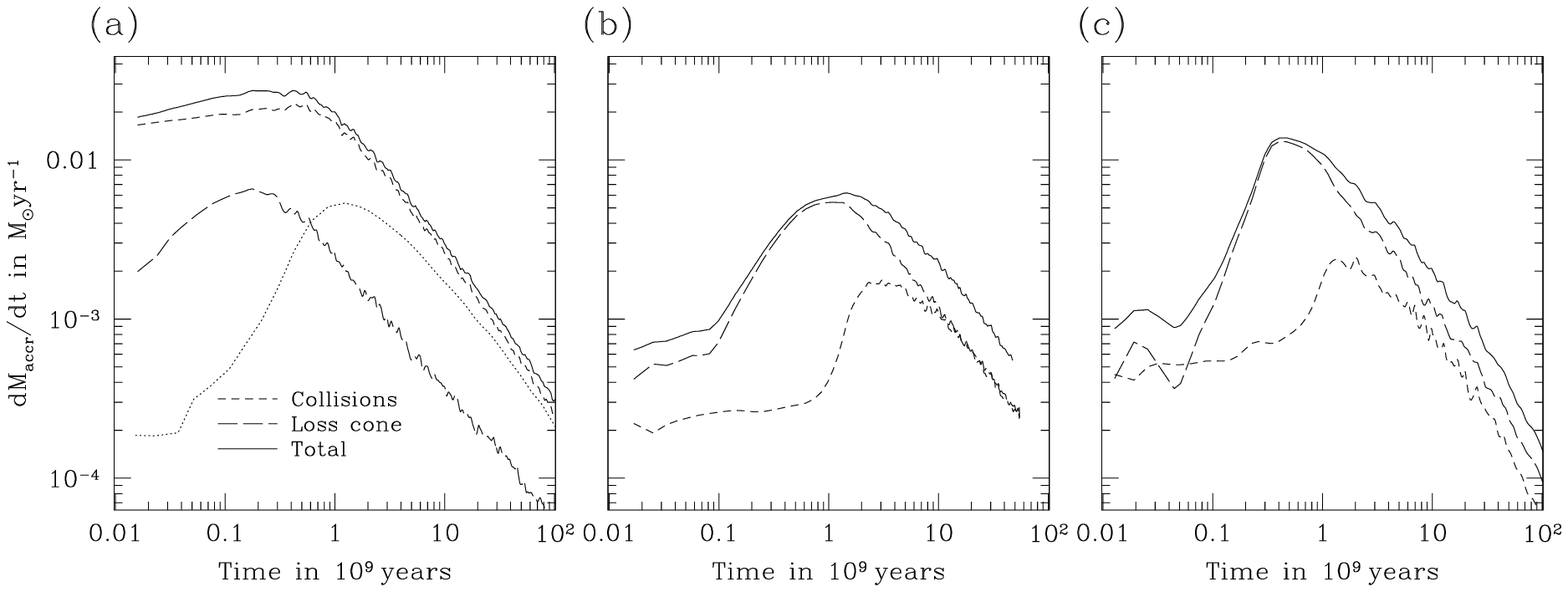}
  \caption{Accretion rate on the central BH. We assume instantaneous and
    complete accretion of the gas released in collisions (short
    dashes) or tidal disruptions (long dashes, including horizon
    crossing stars). {\bf Panel (a)}: Completely disruptive collisions
    (as in DS83).  The dotted line shows a simulation without
    collisions. {\bf Panel (b)}: SPH-generated prescriptions for
    collisions. As in panel (a), all stars have initially
    1\,$M_\odot$. {\bf Panel (c)}: Same as (b) but with mass spectrum
    $\mathrm{d}N/\mathrm{M_\ast}\propto M_\ast^{-2.35}$ over
    $0.35$--$17.3\,M_\odot$ ($\langle M_\ast \rangle = 1\,M_\odot$).
    }
  \label{fig:dmdt}
\end{figure}

\begin{figure}[t!]
  \plotone{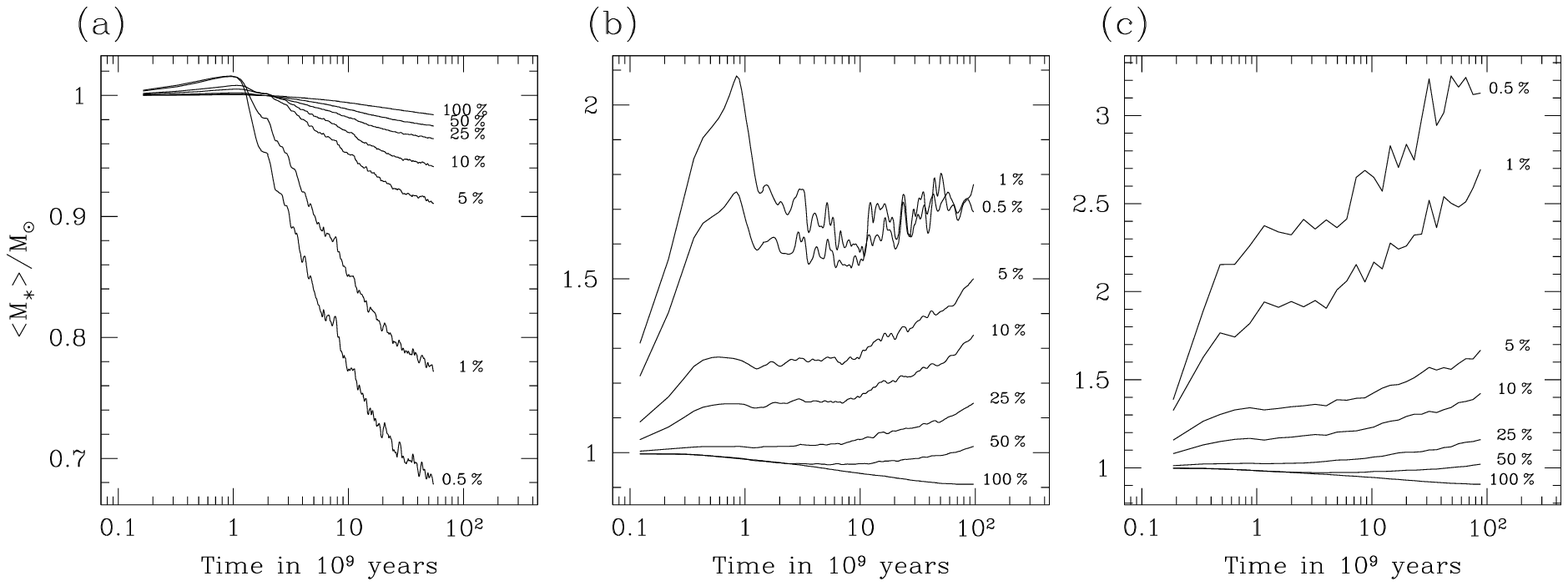}
  \caption{Evolution of the spatial distribution of stellar masses.
    We show the stellar mass averaged over Lagrangian spheres
    containing 0.5 to 100\,\% of the total cluster mass. {\bf Panel
      (a)}: Single mass model, same as
    in Fig.~\ref{fig:dmdt}b. {\bf Panel (b)}: Extended mass spectrum, same
    as in Fig.~\ref{fig:dmdt}c. {\bf Panel (c)}: Same as (b),
    without collisions.}
  \label{fig:segr}
\end{figure}

To assess the influence of realistically treated collisions, we
simulated nuclei models similar to those investigated by Duncan \&
Shapiro (1983, DS83). These are $W_0=8$ King models made of $3.6\times
10^8$ 1\,$M_\odot$ stars with an initial central density of $\sim
7\times 10^7\,\mathrm{pc}^{-3}$ and a $5\times 10^4\,M_\odot$ seed
central black hole which is allowed to grow by accreting gas released
in stellar collisions and tidal disruptions. Some of our results for
these systems are depicted in Figs.~\ref{fig:dmdt} and \ref{fig:segr}.
When we treat collisions the same way as DS83 did, either by
neglecting them completely or, on the contrary, by assuming that every
collision leads to complete disruption of both stars, we get results
in good agreement with those of DS83. In particular, collisions
completely dominate the BH's feeding. However, when realistic
(SPH-generated) prescriptions are used for the collisional outcome,
most events, being grazing encounters, turn out to cause very limited
mass loss and the growth rate is only slightly increased as compared
to the simulation where collisions are neglected. Also, a steep inner
density cusp forms at late evolution stages with a slope closer to
$\alpha=1.75$ than to $\alpha=0.5$.  An intriguing feature of
Fig.~\ref{fig:dmdt} is that, at late times, although the relative
contribution of collisions and tidal disruptions is very different from
one simulation to another, the total accretion rate is nearly the same
in all cases, as if the cluster adjusts itself to ensure a given mass
consumption rate, regardless of the details of disruptive processes.
Further investigations should cast more light on this behaviour,
reminiscent of the binary-driven post core-collapse evolution of
globular clusters.

Fig.~\ref{fig:segr} illustrates how the stellar mass spectrum changes
with time and position in the cluster as a result of mass segregation,
collisions and tidal disruptions. Here the effect of collisions is
more obvious than on the overall stellar structure. When no initial
mass spectrum is present, they lead to a clear decrease of the average 
stellar mass ($\langle
M_\ast \rangle$) in the center-most regions (after a short initial
merging phase). With an extended initial mass spectrum, mass
segregation leads to a quick rise of the central values of
$\langle M_\ast \rangle$. At later stages, as central relative
velocities close to the growing BH get higher and higher, collisions
are more and more effective and prevent any further rise of $\langle
M_\ast \rangle$.

\begin{references}
  
  \reference Duncan, M. J., \& Shapiro, S. L. 1983, \apj, 268, 565

  \reference Freitag, M., \& Benz, W. 2000a, ``A new Monte Carlo code
  for star cluster simulations'', in preparation
  
  \reference Freitag, M., \& Benz, W. 2000b, ``A comprehensive set of
  collision simulations between main sequence stars'', in preparation
  
  \reference H\'enon, M. 1973, in 3rd Advanced Course of the SSAA,
  Dynamical structure and evolution of stellar systems, ed. L.
  Martinet \& M. Mayor, 183

  \reference Spitzer, L. , Jr., \& Saslaw, W. C. 1966, \apj, 143, 400

\end{references}
\end{document}